# 22-Step Collisions for SHA-2

Somitra Kumar Sanadhya[*] and Palash Sarkar

Applied Statistics Unit,
Indian Statistical Institute,
203, B.T. Road, Kolkata,
India 700108.
somitra_r@isical.ac.in, palash@isical.ac.in

$8^{th}$ March, 2008

**Abstract.** In this note, we provide the first 22-step collisions for SHA-256 and SHA-512. Detailed technique of generating these collisions will be provided in the next revision of this note.

## 1 Introduction

SHA-256 and SHA-512 are the next generation hash functions designed and standardized by NIST in 2002 [1]. In this note, we provide message pairs colliding for 22-step SHA-256 and 22-step SHA-512. This is the first attack on 22-step SHA-2 family. The success probability of our attack is around $2^{-5}$ in average case and around $2^{-9}$ in the worst case. Both these probability figures are experimental. Details of the attack will be provided in the next revision of this note.

## 2 Message pairs colliding for 22-step SHA-2

**Table 1.** Colliding message pair for 22-step SHA-256 with standard IV. These message pairs follow the differential path given in Table 3.

| $W_1$ | 0-7 | a0263fa5 | 707425fb | 618cd8d2 | 7d58f729 | 1eb9a964 | 19f88f1c | 34e35071 | f28d40e3 |
|---|---|---|---|---|---|---|---|---|---|
|  | 8-15 | b43e29b8 | 1871a949 | e2e01390 | aaf3823e | 8d41a28e | 7f22ee02 | 7c625999 | 183e603f |
| $W_2$ | 0-7 | a0263fa5 | 707425fb | 618cd8d2 | 7d58f729 | 1eb9a964 | 19f88f1c | 34e35071 | f28d40e3 |
|  | 8-15 | b43e29b9 | 1871a948 | defe7410 | aaf5223e | 8d41a28e | 7f22ee02 | 7c625999 | 00000000 |

**Table 2.** Colliding message pair for 22-step SHA-512 with standard IV.

| $W_1$ | 0-3 | 3ffb91948b327337 | 95f3c893b2356b98 | 506c68760abf51e9 | fab877b7eef3aaa2 |
|---|---|---|---|---|---|
|  | 4-7 | 55d5b38ec34340cf | daa006ef3f677afa | a5a01d9f1c67d9c8 | 5b219ee6f447480b |
|  | 8-11 | 52af39ff1ecfb48e | 5cff9ae5d4d60a40 | db6c1a412c9b4d4d | aaf3823c2a004b1f |
|  | 12-15 | 8d41a28b0d847693 | 7f212e01c4e96937 | 7eeeca5c84ba3bda | 1acad103aa814e0e |
| $W_2$ | 0-3 | 3ffb91948b327337 | 95f3c893b2356b98 | 506c68760abf51e9 | fab877b7eef3aaa2 |
|  | 4-7 | 55d5b38ec34340cf | daa006ef3f677afa | a5a01d9f1c67d9c8 | 5b219ee6f447480b |
|  | 8-11 | 52af39ff1ecfb48f | 5cff9ae5d4d60a3f | db687a412d1b4d65 | aaf3623c2a004b07 |
|  | 12-15 | 8d41a28b0d847693 | 7f212e01c4e96937 | 7eeeca5c84ba3bda | 0000000000000000 |

---

[*] This author is supported by the Ministry of Information Technology, Govt. of India.

**Table 3.** Differential path followed by the message pairs given in Table 1. The differential path for the message pair of Table 2 is different but similar looking. If the register values for the first message $W_1$ are denoted by $\{a_i, b_i, \ldots h_i\}$ and those for the second message $W_2$ are denoted by $\{a'_i, b'_i, \ldots h'_i\}$, then $\delta X$ stands for $X' - X$, where $X$ could be any register value. The 22 steps are indexed from 0 to 21.

| Step | $\delta a_i$ | $\delta b_i$ | $\delta c_i$ | $\delta d_i$ | $\delta e_i$ | $\delta f_i$ | $\delta g_i$ | $\delta h_i$ |
|---|---|---|---|---|---|---|---|---|
| 0 | 0 | 0 | 0 | 0 | 0 | 0 | 0 | 0 |
| 1 | 0 | 0 | 0 | 0 | 0 | 0 | 0 | 0 |
| 2 | 0 | 0 | 0 | 0 | 0 | 0 | 0 | 0 |
| 3 | 0 | 0 | 0 | 0 | 0 | 0 | 0 | 0 |
| 4 | 0 | 0 | 0 | 0 | 0 | 0 | 0 | 0 |
| 5 | 0 | 0 | 0 | 0 | 0 | 0 | 0 | 0 |
| 6 | 0 | 0 | 0 | 0 | 0 | 0 | 0 | 0 |
| 7 | 0 | 0 | 0 | 0 | 0 | 0 | 0 | 0 |
| 8 | 00000001 | 0 | 0 | 0 | 00000001 | 0 | 0 | 0 |
| 9 | 0 | 00000001 | 0 | 0 | ffffffff | 00000001 | 0 | 0 |
| 10 | 0 | 0 | 00000001 | 0 | ffffffff | ffffffff | 00000001 | 0 |
| 11 | 0 | 0 | 0 | 00000001 | 0 | ffffffff | ffffffff | 00000001 |
| 12 | 0 | 0 | 0 | 0 | 00000001 | 0 | ffffffff | ffffffff |
| 13 | 0 | 0 | 0 | 0 | 0 | 00000001 | 0 | ffffffff |
| 14 | 0 | 0 | 0 | 0 | 0 | 0 | 00000001 | 0 |
| 15 | 0 | 0 | 0 | 0 | 0 | 0 | 0 | 00000001 |
| 16 | 0 | 0 | 0 | 0 | 0 | 0 | 0 | 0 |
| 17 | 0 | 0 | 0 | 0 | 0 | 0 | 0 | 0 |
| 18 | 0 | 0 | 0 | 0 | 0 | 0 | 0 | 0 |
| 19 | 0 | 0 | 0 | 0 | 0 | 0 | 0 | 0 |
| 20 | 0 | 0 | 0 | 0 | 0 | 0 | 0 | 0 |
| 21 | 0 | 0 | 0 | 0 | 0 | 0 | 0 | 0 |